\documentclass[%
 reprint,
bibnotes,
 amsmath,amssymb,
 aps, jmp,
floatfix,
]{revtex4-2}
\usepackage{amssymb}
\usepackage{enumitem}
\usepackage{graphicx}
\usepackage{dcolumn}
\usepackage{bm}

\usepackage{blindtext}
\usepackage{hyperref}
\usepackage{lipsum}
\usepackage{mwe}
\usepackage{hyperref}
\usepackage[normalem]{ulem} 
\usepackage[dvipsnames, usenames]{xcolor}
\usepackage{enumitem}
\usepackage{cleveref}

\newlist{algorithm}{enumerate}{4}

\setlist[algorithm, 1]
{label=\Alph{algorithmi},
leftmargin=\parindent,
rightmargin=10pt
}

\setlist[algorithm, 2]
{label=\arabic{algorithmii}, 
leftmargin=15pt,
rightmargin=15pt}

\setlist[algorithm, 3] 
{label=\arabic{algorithmii}.\arabic{algorithmiii},
leftmargin=20pt,
rightmargin=20pt}

\setlist[algorithm, 4] 
{label=\roman{algorithmiv}),
leftmargin=25pt,
rightmargin=25pt}

\begin{document}

\preprint{APS/123-QED}

\title{Automated determination of the end time of junk radiation in binary black hole simulations
}

\author{Isabella G. Pretto$^{1,}$$^{*}$, Mark A. Scheel$^{1}$, and Saul A. Teukolsky$^{1,2}$\\
        \small $^{1}$Theoretical Astrophysics 350-17, California Institute of Technology, Pasadena, California 91125, USA\\
        \small $^{2}$Cornell Center for Astrophysics and Planetary Science, Cornell University, Ithaca, New York 14853, USA}

\begin{abstract}
When numerically solving Einstein’s equations for the evolution
of binary black holes, physical imperfections in the initial
data manifest as a transient, high-frequency pulse of “junk
radiation.” This unphysical signal must be removed before the
waveform can be used.
Improvements in the efficiency of numerical simulations now allow
waveform catalogs containing thousands of waveforms to be produced. Thus, an automated procedure for identifying junk radiation is required.
To this end, we present a new algorithm based
on the empirical mode decomposition (EMD) from the Hilbert-Huang
transform. This approach allows us to isolate and measure the
high-frequency oscillations present in the measured
irreducible masses of the black holes.
The decay of these oscillations allows us to estimate the time from
which the junk radiation can be ignored. To make this procedure
more precise, we propose three distinct threshold criteria that
specify how small the contribution of junk radiation has to be
before it can be considered negligible. We apply this algorithm
to 3403 BBH simulations from the SXS catalog to find appropriate values for the thresholds in the three criteria. We find
that this approach yields reliable decay time estimates, i.e.,
when to consider the simulation physical, for $>98.6\%$ of the
simulations studied. This demonstrates the efficacy of the EMD as
a suitable tool to automatically isolate and characterize junk
radiation in the simulation of binary black hole systems.

\end{abstract}

\maketitle


\section{\label{sec:intro}Introduction}

Colliding binary black holes are among the strongest sources of gravitational waves. Future detectors, such as the Cosmic Explorer and the Einstein Telescope,  will be able to measure their emission with unprecedentedly high detail, making the development of increasingly accurate templates all the more necessary. 

To produce the complete gravitational wave signal, analytical approximations are used to evolve the binary system from large separations until close to merger, when the black holes approach the ultra-relativistic regime and a numerical solution becomes necessary. 

The numerical evolution of Einstein’s field equations is based on the decomposition of spacetime into three-dimensional spacelike hypersurfaces of constant time. In this formalism, the initial data that must be defined for the initial hypersurface consist of the spatial metric $g_{ij}$ and the extrinsic curvature $K_{ij}$. These should satisfy the Hamiltonian  and momentum constraints,
\begin{align}
 \label{eq:hamiltonian}
    G_{nn} &= 0,\\
    G_{nj} &= 0, 
    \label{eq:momentum}
\end{align}
where the subscripts $n$ and $j$ refer to components normal and tangential to the initial slice \cite{Pfeiffer, York}.

Initial data that are astrophysically relevant must serve as a snapshot of the system that has been evolving since $t=-\infty$. However, even with  complete knowledge of this past history, it is not known how to construct a set of initial data that perfectly characterizes the binary in quasi-equilibrium. Features of the system, such as tidal deformations and deviations from conformal flatness, are necessarily simplified, leading to perturbations in the true geometry. As the metric relaxes to equilibrium, this mismatch manifests itself as an initial high-frequency and high-amplitude pulse of ``junk radiation."

To extract the physically relevant gravitational signal and calculate the initial parameters, one must first evolve the system with imperfect initial data until junk radiation leaves the computational domain. Adequately removing this spurious burst is therefore necessary to produce waveforms that are free from such effects, which could lead to biased analyses and hide physical information.

Different approaches have been employed to mitigate the impact of junk radiation, either by constructing more astrophysically accurate initial data or by investigating its long-term effects. These include, for example, departure from the conformal flatness assumption \cite{Lovelace, Johnson-McDaniel,kelly,slinker}, implementation of realistic tidal deformations \cite{Chu}, and specific choices of constraint damping parameters \cite{ma} and coordinate systems \cite{zhang}. They also involve considerations as to how junk radiation, as it propagates away, may lead to constraint violations \cite{Higginbotham} and  affect the input parameters, such as mass, spin and eccentricity, and merger dynamics \cite{bode}. 

Despite these efforts to reduce the effects of junk radiation, the initial transient remains non-negligible. In addition to contaminating the waveform, junk radiation also affects the masses, spins and orbital parameters of the binary, such as eccentricity.  Therefore, the common practice is to ignore the portion of the simulation before some time $t=t_\mathrm{junk}$, when the junk radiation is deemed to be sufficiently small, and to redefine the ``initial'' values of masses, spins, and orbital parameters to be their values at $t=t_\mathrm{junk}$ rather than at $t=0$.

Using the waveform itself to determine $t_\mathrm{junk}$ would mean evolving trial initial data long enough to extract the waveform at some
region sufficiently far from the source. Instead, one can use data from the black holes themselves, such as their trajectories or irreducible masses. 
Currently, in the numerical relativity Spectral Einstein Code (SpEC)  \cite{spec}, {$t_\mathrm{junk}$} is calculated based on the decrease of the average standard deviation of the irreducible mass of individual black holes for sliding time windows, as described in \cite{sxs}. The irreducible mass is defined as
\begin{equation}\label{eqn:areal mass}
M_\mathrm{irr}\equiv \sqrt{\frac{A}{16\pi}}= \sqrt{\frac{1}{16\pi} \int_H dA},
\end{equation}
where $H$ indicates the apparent horizon within the current constant-time hypersurface \cite{sxs}. As junk radiation propagates away, the running average decreases. The decay time is then identified as the earliest time at which the average stops decreasing \cite{sxs}.

However, this procedure tends to underestimate the junk radiation timescale, often requiring human intervention to ensure that the processed waveform is indeed clean of these perturbations. For a catalog with thousands of simulations, such as the Simulating eXtreme Spacetime (SXS) repository \cite{sxs}, it would be preferable to have a fully automated algorithm.

In this work, we present a new algorithm that estimates  {$t_\mathrm{junk}$} based on the empirical mode decomposition (EMD) from the Hilbert-Huang transform (HHT). Henceforth, we will refer to this method as the `EMD' algorithm, and to the standard-deviation-based implementation described in \cite{sxs} as the `SDV' algorithm.

This paper is structured as follows. In the next section, we introduce the HHT and show how the first component of the EMD can be used to track the evolution of junk radiation. In Sec.~\ref{sec:num impl}, we describe, step-by-step, the algorithm that identifies the end time of junk radiation. In Sec.~\ref{sec:results}, we describe the results obtained by applying the algorithm to 3403 BBH simulations from the SXS catalog \cite{sxscatalog}, and identify four categories of waveforms that are characterized by specific threshold criteria. Based on these observations, we make precise assumptions regarding the expected duration of junk radiation that allow the algorithm to apply the most appropriate of these categories to a given simulation. We also discuss special cases that require variations of the default values found. In Sec.~\ref{sec:conclusion}, we summarize our findings and discuss potential applications of the HHT-EMD approach to numerical simulations.

\section{\label{sec:level1}The Hilbert-Huang Transform}

Humans can visually distinguish junk radiation from the physical
portion of the waveform because junk radiation contains
distinctive
high-amplitude and high-frequency oscillations that happen early in
the evolution. These properties suggest a time-frequency description
of the signal as a means to track the decay of junk radiation as
it leaves the computational domain.

Traditional data analysis methods, however, are defined for time series that are linear and/or stationary. Moreover, many spectral decomposition methods, such as the Fourier and the wavelet transforms, are subject to the Heisenberg uncertainty principle, which states that a continuous time signal cannot be arbitrarily accurate in both time and frequency. Specifically, for any $x(t) \in \text{L}^{2}(\mathbb{R})$,
\begin{equation}
    \Delta^{2}_{t}\Delta^{2}_{\omega}\geq \frac{1}{4} ,
\end{equation}
where $\Delta^{2}_{t}$ and $\Delta^{2}_{\omega}$ represent the spread of $x(t)$ in the time and frequency domains \cite{Reza}.

If the natural understanding of junk radiation is in terms of its spectral properties, these restrictions call for an adaptive method. That is, instead of using an a priori basis in which the time-frequency resolution is subject to the uncertainty principle, we need a basis that, derived from the data themselves, can lead to an alternative definition of frequency. 

The HHT is an adaptive method that serves this purpose. It decomposes a signal into a finite number of components of decreasing frequency ranges, which allows for the localization of nonlinear features. It is a combination of the Hilbert spectral analysis, which provides the required alternative definition of instantaneous frequency, and the empirical mode decomposition. The goal of the EMD is to decompose any time series into intrinsic mode functions (IMFs) to which the Hilbert transform can be applied, leading to a description of the data in terms of physically meaningful instantaneous frequencies \cite{Huang1}.

The HHT is a highly efficient process that has found widespread application in the analysis of scientific data, from feature extraction to signal denoising. It has been used, for example, to investigate the characteristics of sources of highly nonlinear seismic waves \cite{zhang}, estimate the decay timescale of and energy carried by the high-frequency component of a tremor transient \cite{bowman}, and identify hidden cycles associated with glacier-interglacier changes in ice cores \cite{Reza}. Examples that demonstrate the power of the HHT in uncovering patterns hidden in the data abound, motivating its use in the isolation and characterization of gravitational bursts in BBH simulations. 

\subsection{The Hilbert transform}

Let $x(t)$ be an element of a Lebesgue space. Its Hilbert transform $y(t)$ is defined as
\begin{equation}
y(t) = \frac{1}{\pi}P \int_{-\infty}^{\infty} \frac{x}{(t-\tau)}d\tau        , 
\end{equation}
where $P$ stands for the Cauchy principal value of the integral \cite{huang2}. 

One can then define a complex signal whose imaginary part is the Hilbert transform of its real part,
\begin{equation}
z(t) = x(t)+iy(t) = a(t)e^{i\theta(t)},
\end{equation}
where 
\begin{equation}
\begin{split}
a(t) &= (x^{2} + y^{2})^{1/2},\\
\theta(t) &= \tan^{-1}\left(\frac{y}{x}\right)
\end{split}
\end{equation}
correspond to the instantaneous amplitude and phase function. The instantaneous frequency follows from these definitions as
\begin{equation}
\omega(t) = \frac{d\theta}{dt}. 
\end{equation}

However, this definition of the instantaneous frequency, although mathematically precise, may not reflect the actual fluctuations present in the original function. For the Hilbert transform to work properly, the signal must be a purely oscillatory or monocomponent function with a zero reference level--- a condition that is rarely satisfied by physical data \cite{Huang1}. 

It is shown in \cite{huang2} that, to circumvent this challenge, an arbitrary signal can be decomposed into IMFs. These objects satisfy the conditions of the Hilbert transform and can lead to a physically meaningful instantaneous frequency. 

\subsection{Empirical mode decomposition}\label{emd}

The EMD is the procedure by which a signal is resolved into IMFs. These are defined such that (1) the difference between the number of extrema and zero crossings is at most one, and (2) the mean value of the upper and lower envelopes obtained from interpolating across the local maxima and minima is close to zero at any data point \cite{huang2}.

As described in \cite{huang2}, IMFs can be obtained from the input signal $x(t)$ according to the following procedure:
\begin{enumerate}
\item
Identify the local maxima and minima of $x(t)$. \label{procedure_step1}
\item
With a cubic spline, interpolate between the local maxima and minima, forming the upper envelope  $e_\mathrm{max}(t)$ and the lower envelope $e_\mathrm{min}(t)$.
\item Compute the mean $m(t)= (e_\mathrm{max}(t)+e_\mathrm{min}(t))/2$ of the envelopes.
\item Extract $h(t)=x(t)-m(t)$ from the data. This is a tentative IMF, which we call a protomode.
\item Replace $x(t)$ by $m(t)$ and repeat the previous steps. \label{procedure_step5}
\end{enumerate}

In practice, to find the $i^{th}$ IMF, the procedure has to be refined by a sifting process until the protomode satisfies the definition of an IMF. Specifically, it must satisfy the zero-mean condition as fixed by some stopping criterion that specifies what ``sufficiently close to zero'' means. 

In this work, we use the threshold method introduced in
\cite{rilling}. This criterion is based on the definition of the
mode amplitude
$A_\mathrm{mode}(t)\equiv[e_\mathrm{max}(t)-e_\mathrm{min}(t)]/2$
and the evaluation function $\sigma(t)\equiv  |m(t)/A_\mathrm{mode}(t)|$. In
each iteration, to determine whether a protomode satisfies
the definition of an IMF, the sifting continues until
$\sigma(t)<\theta_{1}$ for some fraction ($1-\alpha$) of the total
signal time, and $\sigma(t)<\theta_{2}$ for the remaining portion
\cite{rilling}. Here, $\alpha$, $\theta_{1}$ and $\theta_{2}$
are parameters chosen by the user.

Satisfying these conditions amounts to iterating steps ~\ref{procedure_step1}--~\ref{procedure_step5} on the protomode $h_{k-1}(t)$, where the index $k$ represents the iteration number, until $h_{k}(t)$ can be considered an IMF. Once this is achieved, we define the first IMF as $c_{1} \equiv h_{k}(t)$. We then calculate the corresponding residual $r_{1}(t)=x(t)-c_{1}(t)$, which contains longer period oscillations, and proceed to step ~\ref{procedure_step5}, treating $r_{1}(t)$ as the initial data. 

The procedure outlined above is then repeated until the residual becomes a function that is monotonic or contains only one extremum. By combining the IMFs $c_{j}$ thus computed and the final, irreducible residual, we obtain the decomposition of the data into components of increasing periods,
\begin{equation}
    x(t) = \sum_{j=1}^{n} c_{j}(t)\ + r_{n}(t). 
\end{equation}

\begin{figure*}
         \includegraphics[width=0.8\textwidth]{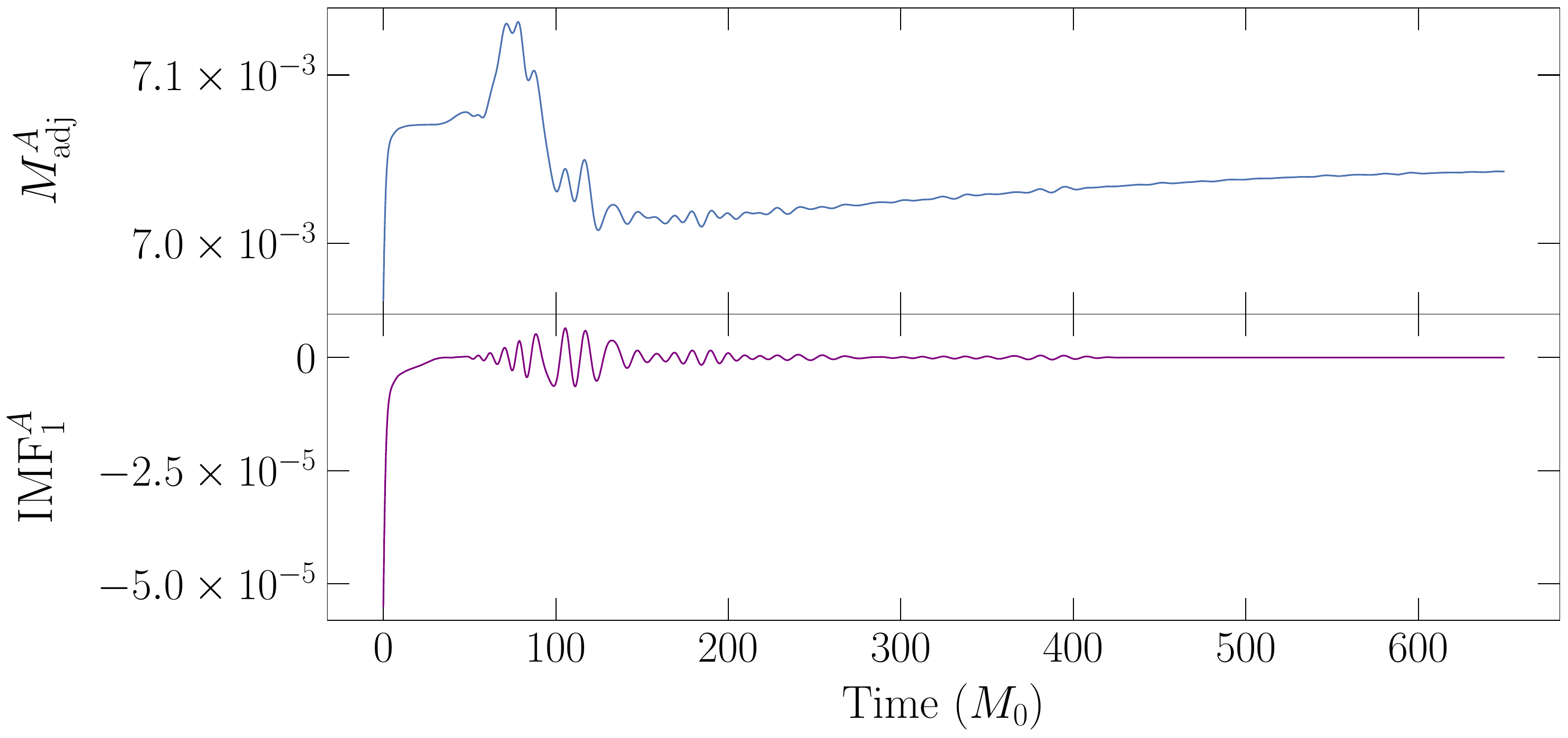}
         \caption{The irreducible mass of the primary black hole from
         SXS:BBH:1465 is decomposed using the EMD.
         The top panel (blue) illustrates the adjusted mass of the black hole, $M_\mathrm{adj}(t)$, from which a mass offset  $M_\mathrm{offset} = 0.56$ is subtracted
         for plotting purposes. The bottom panel (purple) shows the
         first IMF in terms of $M_{0}$. The first mode effectively
         captures the oscillatory pattern from the upper plot
         that characterizes junk radiation, whose effects become
         negligible after $t\sim 450M_{0}$.}
         \label{fig:fig1}
\end{figure*}

 Fig.~\ref{fig:fig1} shows the application of the EMD to the
irreducible mass of the primary black hole from SXS:BBH:1465. The parameters associated with this and the other simulations analyzed in this paper are displayed in Table~\ref{tab:parameters}. To highlight the features associated with junk radiation, we display only $t<700M_{0}$, where $M_{0}$ denotes the sum of the two Christodoulou masses at $t = 0$. To allow more details to be visible in the figures that follow, we also define the adjusted mass,
 \begin{equation}
 M_\mathrm{adj} = \frac{M_\mathrm{irr} -M_\mathrm{offset}}{M_{0}},
 \end{equation}
 where $M_\mathrm{irr}$ is the irreducible mass from which we subtract a constant $M_\mathrm{offset}$. Since all masses are expressed in units of $M_{0}$, $M_\mathrm{adj}$ is dimensionless.
 
 In the top panel of Fig.~\ref{fig:fig1}, the time series for the adjusted mass of one of the black holes is plotted in blue. To show the fluctuations in the dataset, which are of order $10^{-5}$, we set $M_\mathrm{offset} = 0.567 M_{0}$. We start by noting that junk radiation appears as the high-frequency oscillatory pattern at the start of the simulation that slowly fades away, leaving a signal that is approximately constant after $t\sim450M_{0}$. 
 
 The bottom panel displays the first IMF in purple. Its amplitude is also of order $\sim 10^{-5}$, which is consistent with the fluctuations observed in the irreducible mass due to junk radiation. We further notice that the IMF effectively tracks the evolution of junk radiation shown in the upper panel. It records oscillations of decreasing amplitude, resulting in a signal that remains unchanged after $t\sim 450M_{0}$. Since, by definition, IMFs have a zero mean, these features suggest that the amplitude of the first mode can be used as a way to characterize and measure the contribution of junk radiation to the data. In the following figures, the data for the adjusted mass and first IMF will also be represented in blue and purple.

\section{Algorithm}\label{sec:num impl}

In this section, we systematically outline the junk-determining procedure. 

We note that the algorithm was trained using the irreducible mass of the black holes from the binary. Consequently, the numerical values quoted may change if a different dataset is used. We further describe how the irreducible mass was processed in Sec.~\ref{sec:results}. The algorithm is somewhat complicated
because there are several special cases that need to be handled. Four different
filters are introduced. The rationale for the numerical values chosen for the four filters described below is further justified
in the corresponding sections (Secs.~\ref{crit1} - ~\ref{results_filter4}). 

In what follows, $t_{\mathrm{junk},i}$, where $i \in \{A,B\}$, refers to the time junk radiation ends in the evolution of the irreducible mass of one of the black holes. Note that `A' identifies the black hole with the larger mass. Given this dataset, proceed as follows:

\begin{algorithm}
\item \textbf{Reduce dataset with junk radiation}
\begin{algorithm}
\item Compute the first IMF by performing EMD using the PyHHT Python module.

We note that the first IMF alone is sufficient to track the evolution of junk radiation. Moreover, the number of IMFs into which a signal is decomposed is not known a priori. This is a consequence of the empirical nature of the EMD, which makes limiting the number of terms in the decomposition necessary.\label{stepA1}

\item Truncate the first IMF to eliminate all data
before  $t = t_{0}$ and after $t_\mathrm{end} \equiv t=0.8\times
t_\mathrm{latest}$, where $t_{0}$ is the time of the first
maximum of the IMF and $t_\mathrm{latest}$ is the latest time
in the dataset. An IMF maximum is detected if its amplitude is
greater than that of its two closest neighboring points. That
is, an IMF maximum with index $n$ exists if IMF$[n]>{}$IMF$[n-1]$ and
IMF$[n]>{}$IMF$[n+1]$.\label{stepA2}

The truncation is necessary because, at early and late times, the first IMF contains nearly-vertical features with an amplitude that significantly differs from that of the main portion of junk radiation. At late times, we must remove the high-amplitude maxima that are present in both the waveform and the IMF as the merger time approaches. For the simulations we tested, we found that removing the last $20 \% $ of the simulation time window reliably eliminated these late-time oscillations while preserving the junk radiation content. 
  
This behavior can be observed in Fig.~\ref{fig:fig2}, which shows the irreducible mass and the first IMF for (a) the first $\sim400 M_{0}$ and (b) the last $\sim1000 M_{0}$ of the evolution of the secondary black hole of SXS:BBH:0692. In Fig.~\ref{fig:fig2}a, the orange dashed line identifies the first maximum of the first IMF at $t=30.5 M_{0}$. Notice that the feature at $t<30.5M_{0}$ causes the irreducible mass to increase by $\sim 8\times 10^{-5}$. It is also followed by the main burst of junk radiation, which decays within $t\sim 300 M_{0}$. In Fig.~\ref{fig:fig2}b, the pink dashed line indicates the last maximum of the first IMF at $t = 4789 M_{0}$. In this case, where $t_\mathrm{latest} = 5012 M_{0}$, we find $t_\mathrm{end} = 0.8\times t_\mathrm{latest} = 4009.6 M_{0}$. Therefore, by following Step ~\ref{stepA2} we achieve the desired result of removing the portion of the waveform associated with the merger. The left plot shows that the mass increases by over $2.5\times10^{-4}$ as the merger time approaches. In both cases, the first IMF captures these vertical features with an amplitude that differs from the main portion of junk radiation by over an order of magnitude. Therefore, these early- and late-time regions must be removed to avoid biases in the characterization of junk radiation.    
  
\item Calculate the characteristic amplitude of junk radiation, defined as \texttt{mean\_about\_max}, in the following way:
\begin{algorithm}
\item Calculate $t = t_\mathrm{peak}$, where $t_\mathrm{peak}$ is the time when the maximum amplitude of the first IMF occurs.
\item Fix a time window that isolates the bulk of junk radiation before it starts to significantly decay.

In Fig.~\ref{fig:fig2}a, the main portion of junk radiation can be observed for $30.5<t<250 M_{0}$. 
We found, by considering different numbers of IMF maxima, that a similar region in a general simulation is typically contained within the window [$t_{0}$, $t_\mathrm{peak}$ + $t_{\mathrm{peak}+15}$], where $t_\mathrm{peak+15}$ is the time when the $15^{th}$ maximum after $t_\mathrm{peak}$  occurs.  
\item Define \texttt{mean\_about\_max} as the average of all the IMF maxima contained in this window.
\end{algorithm}\label{A.3} 
\item Truncate the IMF again by removing all data before $t_\mathrm{peak}$.
     
\end{algorithm}\label{partA}

\item \textbf{Calculate} $t_\mathrm{junk}$ \textbf{for a single black hole}
\begin{algorithm}
\item \textbf{Apply filter 1}

\begin{algorithm}
\item If $\texttt{mean\_about\_max}<10^{-9}$, choose $t_{\mathrm{junk},i}$ to be the first time (going forward) when the amplitude of one of the IMF maxima falls below $5\times10^{-10}$ and proceed to Part~\ref{partC}. Else, follow the next steps.

This condition handles the special case where the junk radiation is always small. \label{item5}
\end{algorithm}\label{filter1}

\item \textbf{Apply filter 2}
\begin{algorithm}
\item Starting at the end of the IMF and moving backwards in time, 
      choose the provisional $t_{\mathrm{junk},i}$ as the first time when the IMF exceeds $f\mathrm{\times \texttt{mean\_about\_max}}$, where $f = f_{2} \equiv 0.001$.
        \label{stepB2.1}

\item Assess the goodness of the provisional $t_{\mathrm{junk},i}$ by checking whether
\begin{algorithm}
\item $t_{\mathrm{junk},i} > T$,  or 
\item $t_{\mathrm{junk},i} > F\times t_\mathrm{end}$, 
\end{algorithm}\label{stepB2.2}
where $T = T_{2} \equiv 2000$ and $F = F_{2}\equiv 0.6$.

If both statements are false, store the value of $t_{\mathrm{junk},i}$ and skip to Part~\ref{partC}. Else, $t_{\mathrm{junk},i}$ is considered an overestimate. In this case, proceed to the next step.

Note that the values for $T_{2}$ and $F_{2}$ reflect the user's expectations regarding the timescale of junk radiation. By applying this filter to our simulations, we found that, typically, a value for $t_\mathrm{junk}$ that exceeds either $t=0.6\times t_\mathrm{end}$ or $t=2000$ is indicative of an overestimate. These values function as upper bounds to $t_\mathrm{junk}$. However, if the feature detected with Filter~\ref{filter2} past these thresholds is indeed from junk radiation, we expect it to be identified by the subsequent filters as well.

\end{algorithm}\label{filter2}

\item \textbf{Apply filter 3}
\begin{algorithm}
\item Repeat step B~\ref{stepB2.1}, with $f = f_{3} \equiv 0.01$. \label{step3.1}
\item Repeat step B~\ref{stepB2.2}, with $T = T_{3} \equiv 3000$ and $F = F_{3} \equiv 0.6$. \label{step3.2}

Since the assumption that junk radiation ends before $t = 0.6\times t_\mathrm{end}$ remains valid for most simulations, we set $F_{3}=F_{2}=0.6$. The parameter $T$, however, is modified to accommodate unusually long simulations, where junk radiation can persist past $t=2000 M_{0}$. Our observations indicate that, in these cases, junk radiation is not expected to last longer than $t=T_{3}=3000 M_{0}$. 
It is also worth noting that we do not adopt this threshold in Filter \ref{filter2} because, paired with the low value for the $f_{2}$ parameter, it tends to result in an overestimation of $t_\mathrm{junk}$. The $f$ values used by Filters \ref{filter3} and \ref{filter4} were more appropriate for simulations with persisting junk radiation.
\end{algorithm}\label{filter3}

\item \textbf{Apply filter 4}
\begin{algorithm}
\item Repeat step B.\ref{stepB2.1}, with $f = f_{4} \equiv 0.025$.\label{step4.1}
\item Repeat step B.\ref{stepB2.2}, with $T = T_{4} \equiv 3000$ and $F = F_{4} \equiv 0.6$.
\label{step4.2}
\end{algorithm}\label{filter4}
    
\end{algorithm}\label{partB}

\item \textbf{Calculate the final} $t_{\mathrm{junk}}$
\begin{algorithm}

\item Repeat Parts \ref{partA} and \ref{partB} for the second black hole.

\item With $t_{\mathrm{junk_A}}$ and $t_{\mathrm{junk_B}}$ as the values found in Part \ref{partB} using the time series for the first and second black holes, define the final junk radiation end time as
\begin{algorithm}

\item $t_{\mathrm{junk}} = \mathrm{max}(t_{\mathrm{junk_A}}, t_{\mathrm{junk_B}})$, if the  $t_{\mathrm{junk}}$ of neither black holes is flagged as an overestimate with the parameters from \ref{step4.2}.

\item $t_{\mathrm{junk}} = \mathrm{min}(t_{\mathrm{junk_A}}, t_{\mathrm{junk_B}})$, if the  $t_{\mathrm{junk}}$ of either black holes is flagged as an overestimate with the parameters from \ref{step4.2}. 
We found that only 19 out of 3403 simulations exhibit this behavior. In all of these cases, taking the minimum of $t_{\mathrm{junk_A}}$ and $t_{\mathrm{junk_B}}$ accurately determines the end time of junk radiation. \label{stepC2.2}

\end{algorithm}

\end{algorithm}\label{partC}

\end{algorithm}

\section{Results}\label{sec:results}

In this section, we describe the results obtained by applying the algorithm outlined in Sec.~\ref{sec:num impl} to the binary black hole simulations from \cite{sxscatalog}. 

We start by describing the data that were used to track the evolution of junk radiation. In Sec.~\ref{resultsfirstimf}, we examine the general properties displayed by the first IMF across the simulations analyzed. In Secs.~\ref{crit1}-~\ref{results_filter4}, we illustrate the simulations that correspond to the four filters presented in Step \ref{partB} of Sec.~\ref{sec:num impl}. We also consider in Sec.~\ref{exception}  eccentric, precessing and high-spin systems. We then analyze in Sec.~\ref{limitations} cases in which the algorithm introduced here fails.

\subsection{Selection of data}\label{resultsdata}

In our implementation, we analyze the time series of irreducible masses of both black holes from the binary. The BBH simulations we use were created with SpEC \cite{spec}, where the constraint-satisfying initial data are constructed using the extended conformal thin-sandwich equations (XCTS) \cite{Pfeiffer,York}.

In this formalism, the conformal 3-metric of the initial hypersurface and the trace of its extrinsic curvature are weighted superpositions of the analytic solutions for two single black holes in Kerr-Schild coordinates. The time derivative of the conformal 3-metric, which can also be freely chosen, is set to zero to reflect the quasi-equilibrium condition of the system. The XCTS equations are then solved using an elliptic solver and yield the initial data for the BBH evolution; namely, the spatial metric, extrinsic curvature, lapse and shift \cite{sxs}. The initial data are subsequently evolved using the generalized harmonic formalism \cite{pretorious, Helmut, Garfinkle, Lindbloom}. 

We analyzed a total of 3403 SXS BBH simulations that include eccentric and spin-precessing systems, mass ratios between 1 and 15, and spin magnitudes up to 0.998. Of these, 1894 simulations are from the public SXS catalog, and 1509 are not yet publicly available as of the release of this paper. These currently private simulations will be marked with an asterisk in the remainder of this paper. Unless otherwise specified, the data used in our analyses are generated from the highest resolution run.

To perform the empirical mode decomposition of the irreducible mass, we use the Hilbert-Huang transform implementation available in the Python package PyHHT \mbox{v.~0.0.1}. We start by creating an EMD object using the \texttt{pyhht.emd.EMD} function. With the method \texttt{decompose}, we then calculate the first IMF component. Except where noted, we accept the default values for the optional parameters in the callables used. These are \texttt{threshold\_1}=0.05, \texttt{threshold\_2}=0.5, \texttt{alpha}=0.05, \texttt{ndirs}=4, \texttt{fixe}=0, \texttt{maxiter}=2000, \texttt{fixe\_h}=0, \texttt{n\_imfs}=0, \texttt{nbsym}=2, \texttt{bivariate\_mode}=\textquotesingle bbox\_center\textquotesingle.

\begin{figure*}[tb]
\centering
\includegraphics[width=\linewidth]{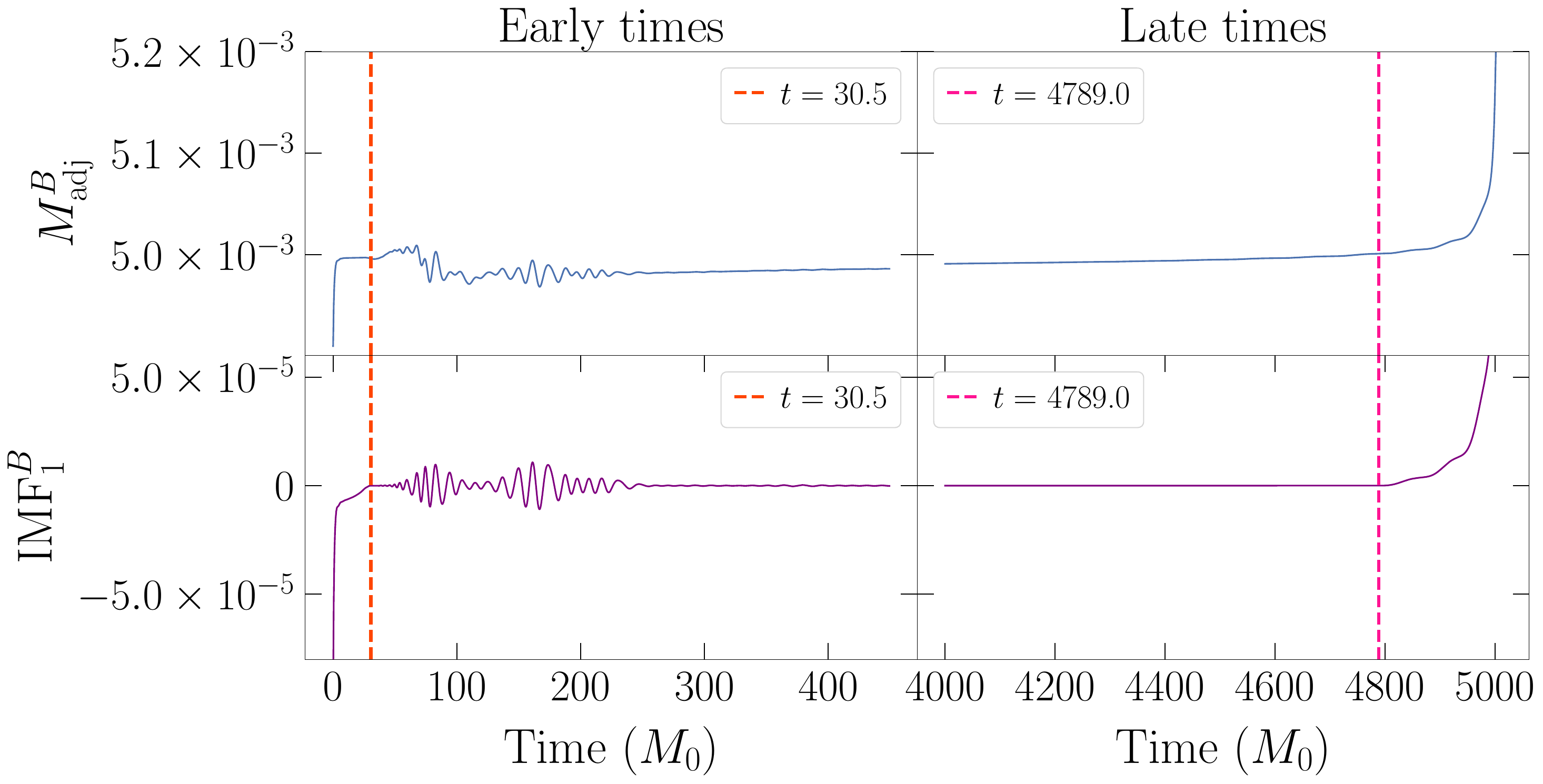}
\caption{The adjusted mass (blue) and first IMF (purple) of the secondary black hole of SXS:BBH:0692 for  $t\le400$ (early times) and $4000\le t\le 5000$ (late times). The mass at early and late times is offset by $0.33 M_{0}$ and $0.335 M_{0}$. The orange and magenta dashed lines indicate the first and last peak of the IMF. The largest variations in the irreducible mass are of order $\sim10^{-5}$, and appear at $t\lesssim 30.5$ and $t\gtrsim 4789$. In the IMF, these correspond to an increase in amplitude by over an order of magnitude. These regions are isolated to the left/right of the first/last IMF maxima and are not relevant for determining $t_\mathrm{junk}$.}
\label{fig:fig2}
\end{figure*}

\subsection{Characterization of the first IMF}\label{resultsfirstimf}

We have shown in Sec.~\ref{emd} that the first IMF adequately tracks the evolution of junk radiation. In this section, we identify the specific properties of junk radiation that are captured by the first IMF and are common across simulations. 

\begin{figure*}[tb]
\centering
\includegraphics[width=\textwidth]{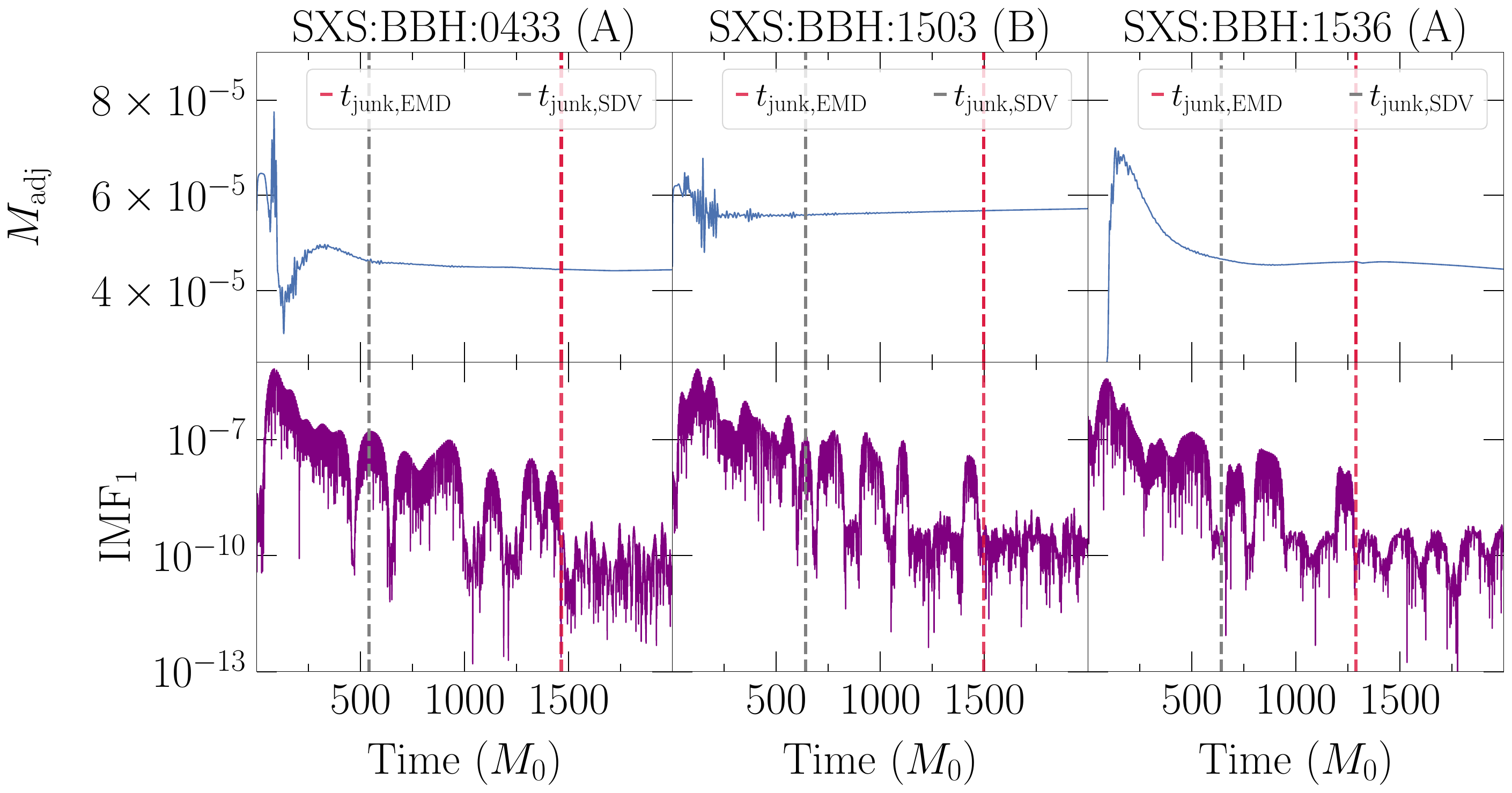}
\caption{The adjusted irreducible mass (top/blue) and the corresponding first IMF (bottom/purple) for the primary (`A') and secondary (`B') black holes from SXS:BBH:0433, SXS:BBH:1503 and SXS:BBH:1536. The irreducible mass in each of these cases is offset by $0.6324 M_{0}$, $0.45918 M_{0}$ and $0.73785 M_{0}$. The values of $t_\mathrm{junk}$ obtained using the SDV- and EMD-based algorithms are shown as $t_\mathrm{junk, SDV}$ (grey/dashed) and $t_\mathrm{junk, EMD}$ (red/dashed) for the various simulations. Junk radiation ($t \lesssim 1500$) has distinct features in the adjusted mass plot of the different simulations. Nonetheless, the IMFs are nearly identical.}\label{fig:fig3}  
\end{figure*}

Fig.~\ref{fig:fig3} shows in the top/bottom panels the time series for the adjusted mass  and the corresponding first IMF for three BBH simulations. The data from the primary and secondary black holes are labeled by `A' and `B'. The red dashed lines represent the value of $t_\mathrm{junk}$ that is returned by the EMD-based algorithm ($t_\mathrm{junk,EMD}$) evaluated at the primary and secondary black holes. The gray dashed line indicates the $t_\mathrm{junk}$ that is returned by the standard-deviation-based algorithm ($t_\mathrm{junk,SDV}$). All figures displayed in the remainder of this paper follow the color and labeling conventions just described. 

We start by noticing that the global behavior of the initial burst of junk radiation in the irreducible mass plots (blue) is significantly different across the simulations considered. In  Fig.~\ref{fig:fig3}a, we observe that the first panel has three distinct extrema with amplitudes $\sim 8 \times 10^{-5}$, $3\times 10^{-5}$ and $5 \times 10^{-5}$, before it stabilizes at $\sim 4.5 \times 10^{-5}$ within $t\sim 1000 M_{0}$. 
The main portion of junk radiation in the corresponding plot of Fig.~\ref{fig:fig3}b has amplitudes in the range  $\sim 3.0-4.5 \times 10^{-5}$. It remains approximately symmetric about the trend as it decays within $t\sim 1500 M_{0}$. In  Fig.~\ref{fig:fig3}c, the amplitude starts close to zero and reaches a maximum at $\sim 7 \times 10^{-5}$, after which it decays within $t\sim 1300 M_{0}$ and stabilizes at $\sim 5 \times 10^{-5}$.

Despite these differences, the corresponding IMFs (purple) display similar features. In the three cases, the amplitude of the first mode abruptly increases at the start of the simulation, reaches a maximum and decays almost exponentially. As it fades away, we notice the appearance of isolated bursts with an average amplitude that tends to decrease with time, after which the signal remains constant at a $\sim 10^{-13}$ numerical floor. These observations suggest that, regardless of the behavior of the underlying trend, the first mode isolates the high-frequency oscillations only, resulting in first IMFs with similar patterns. 

In the three simulations, from left to right and in terms of $M_{0}$, $(t_\mathrm{junk, EMD}, t_\mathrm{junk, SDV})$ is given by (1464.5, 540.0), (1496.5, 640), (1287.5, 640.0). Compared to $t_\mathrm{junk, SDV}$, $t_\mathrm{junk, EMD}$ increases by $924.5 M_{0}$, $856.5 M_{0}$ and $647.5 M_{0}$. As can be seen in the corresponding IMFs, this is because the mode decomposition allows for an efficient isolation of the high-frequency features characteristic of junk radiation. Our goal is to turn these features into an automated algorithm to estimate $t_\mathrm{junk}$. 

An estimate is deemed suitable if it minimizes the presence of junk radiation while preserving most of the physical portion of the signal. In situations where we are confronted with a choice between two competing values, each resulting in a slight overestimation or underestimation of $t_\mathrm{junk}$, our preference is to err on the side of slight overestimation. That is, we would rather sacrifice a small portion of the physically relevant signal than admit persisting junk radiation. 

The algorithm described in Sec.~\ref{sec:num impl} is complicated because it must handle a variety of special cases.  Some
of the steps in that section are designed to treat some of these simulations. For each of the following subsections, we describe a
specific step from Part III~\ref{partB} that determines the final value of $t_\mathrm{junk}$ for some subset of simulations, and we show a representative simulation in this subset. The subsections are labeled according to the corresponding steps in Part III~\ref{partB}.

\subsection{Filter 1}\label{results_filter1}

Filter ~\ref{filter1} is suitable for simulations that contain only
a small amount of junk radiation (see Table~\ref{tab:simulation_ids1}).    

\begin{figure}
\centering
\includegraphics[width=0.5\textwidth]{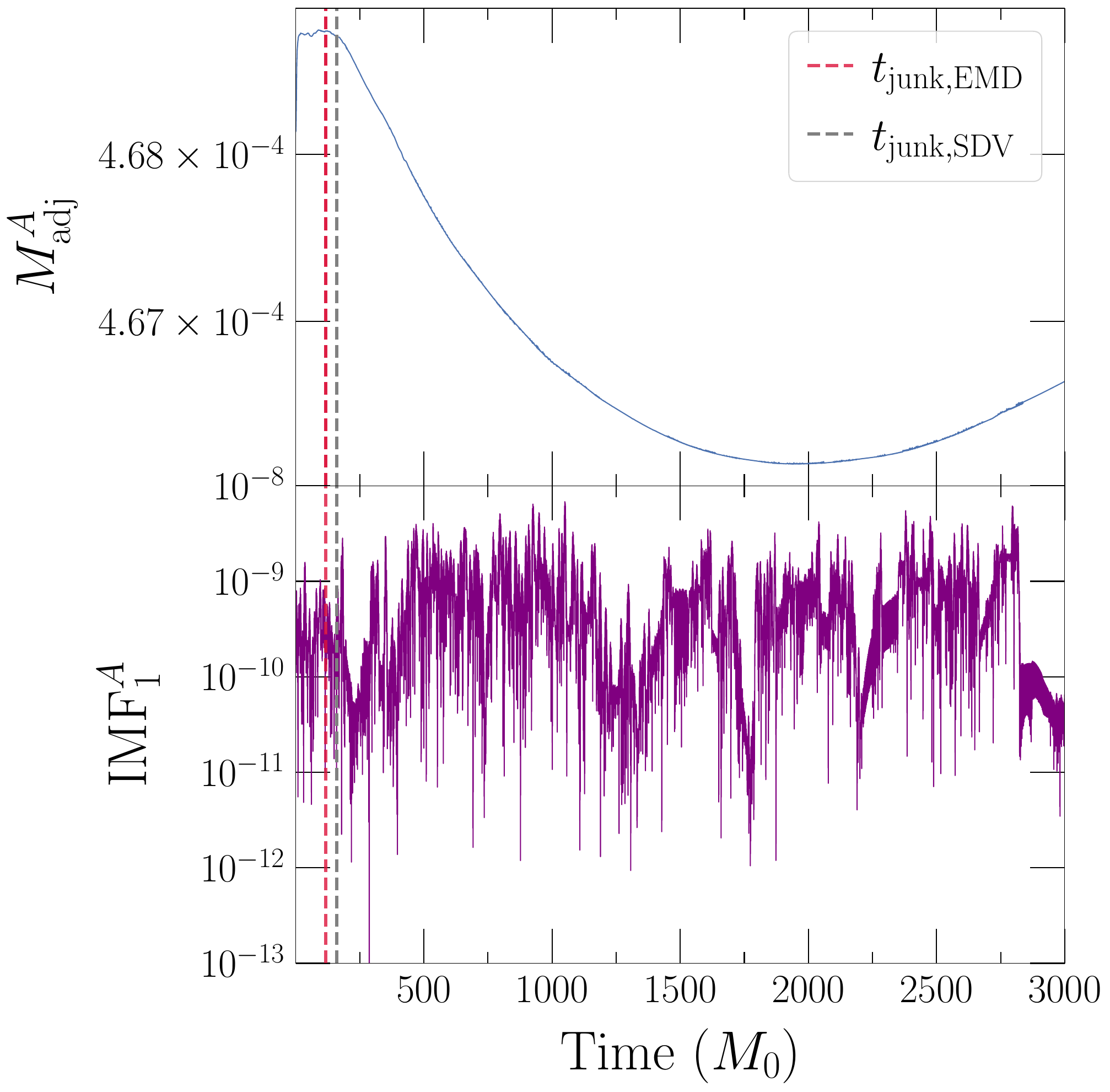}     
 \caption{Filter \ref{filter1} applied to SXS:BBH:1121. The top panel shows in blue the initial burst from junk radiation in the data for the adjusted mass, which has $M_\mathrm{offset}=0.497$. Because the burst decays after only $\sim 160 M_{0}$, the first IMF (bottom/purple) extracts the highest-frequency components of the physical signal itself, which have an average amplitude of $10^{-9}$. The estimate from the EMD- and SDV-based algorithms are shown at $t_\mathrm{junk,EMD} = 116.5 M_{0}$ (red/dashed) and $t_\mathrm{junk,SDV} = 160 M_{0}$ (grey/dashed). }\label{fig:fig7}
\end{figure}

An example is displayed in Fig.~\ref{fig:fig7}, which shows the
adjusted mass and first IMF of SXS:BBH:1121. As can be seen in the
plot for the adjusted mass, junk radiation is already negligible from
the start of the simulation. Correspondingly, the first IMF shows
a valley of amplitude $4.3\times10^{-10}$ at $t_\mathrm{junk_{A},
EMD}=60M_{0}$, followed by oscillations of order $\sim 10^{-9}$
throughout the entire evolution. This behavior is different from that
observed in the simulations shown so far, where the IMF consists
of a series of bursts of decreasing amplitude. This difference
occurs because, in the absence of a significant portion of the
high-amplitude high-frequency oscillations that are characteristic of
junk radiation, the high-frequency components extracted by the first
IMF are likely numerical artifacts from the physical signal itself.

We found that a similar pattern is followed by every other simulation from this category. This observation is encapsulated in the condition that junk radiation can be neglected when the amplitude of the first IMF is smaller than $5\times10^{-10}$.

In this example, the SDV- and EMD-based estimates are $t_\mathrm{junk,SDV}=160 M_{0}$ and $t_\mathrm{junk,EMD}=116.5 M_{0}$. This is an exceptional case in which the SDV approach outperforms the EMD one. However, given how small the additional oscillations captured by the SDV method are, this difference in output is negligible.

\subsection{Filter 2}\label{crit1}
 
Filter~\ref{filter2} has the most sensitive criteria that can be used to identify junk radiation. We found that the choice of $f = 0.001$ used in Step B~\ref{stepB2.2} corresponds to the lowest threshold value that still enables the identification of junk radiation, and that smaller values consistently lead to an overestimate of $t_\mathrm{junk}$. We also note that the overall performance of the algorithm is sensitive to variations in the value of $f$ of order $\sim10^{-3}$. 

\begin{figure}
\centering
\includegraphics[width=0.5\textwidth]{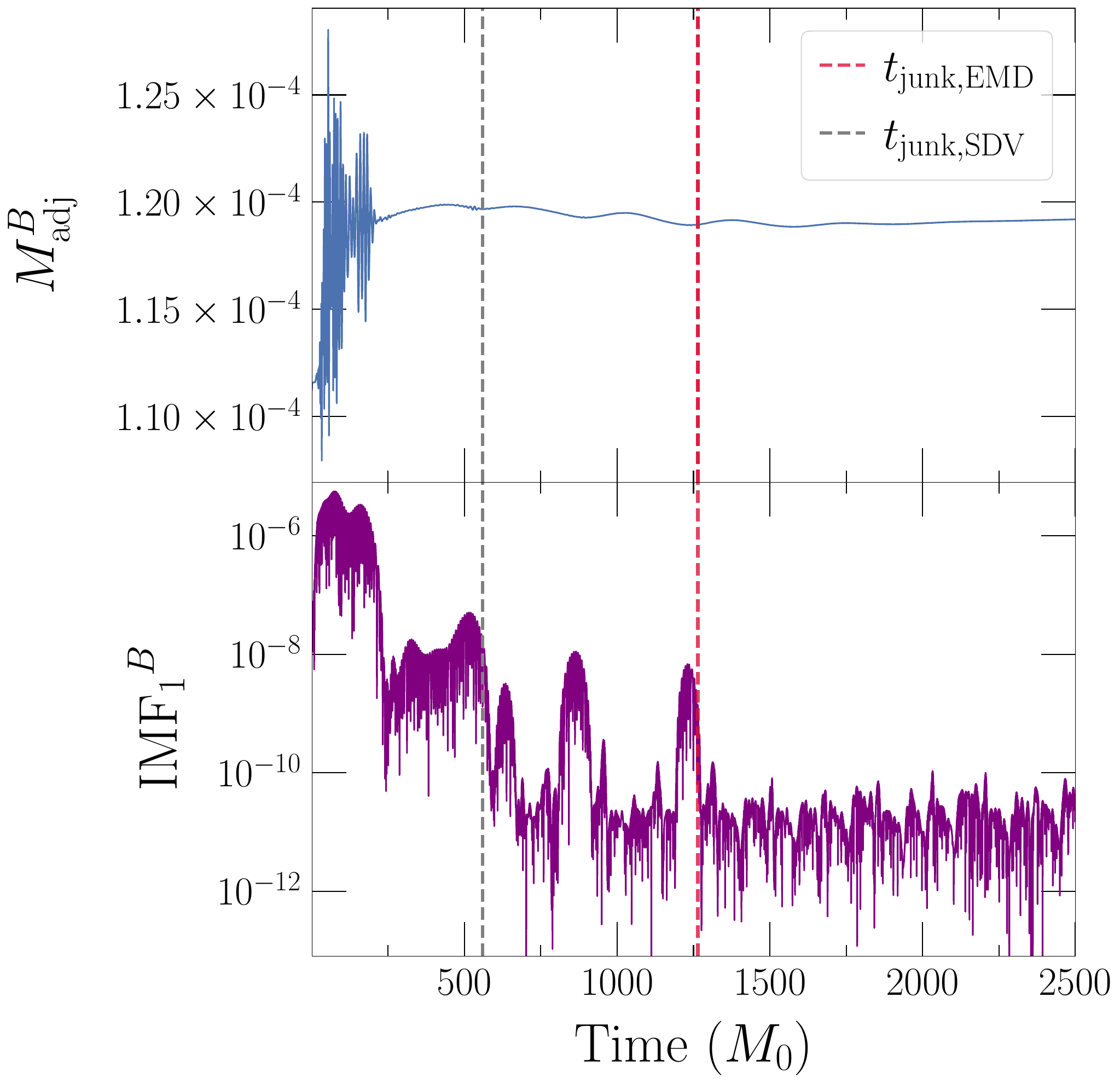}
\label{fig:y equals x}
\caption{The adjusted mass, with $M_\mathrm{offset} = 0.111$, and the first IMF from SXS:BBH:278$1^{*}$. Junk radiation appears after the initial burst as decaying high-frequency fluctuations on a lower-frequency pattern ($560\lesssim t\lesssim 1260$). The output from the SDV and EMD approaches are indicated at $t_\mathrm{junk, SDV} = 560 M_{0}$ and  $t_\mathrm{junk, EMD} = 1264.5 M_{0}$. In the IMF, these late fluctuations appear as bursts of decreasing amplitude.}\label{fig:fig4}
\end{figure}

This step successfully terminates the algorithm primarily for simulations in which junk radiation is associated with a low-frequency tail that decays slowly. This is the case, for example, for SXS:BBH:278$1^{*}$. This scenario is illustrated in Fig.~\ref{fig:fig4}, where the data are displayed according to the prescriptions given for Fig.~\ref{fig:fig3}. We show only the first  $3000 M_{0}$ of data, after which the adjusted mass continues approximately constant up until merger. The SDV and EMD methods estimate the end of junk radiation at $t_\mathrm{junk,SDV}=560 M_{0}$ and $t_\mathrm{junk,EMD}=1264.5 M_{0}$, which are indicated as the grey and red dashed lines. The region that is demarcated by these two lines shows that the EMD algorithm outperforms the SDV one by effectively mapping and isolating the high-frequency fluctuations that ride on the lower-frequency trend for $500\lesssim t\lesssim 2000 M_{0}$. These are featured as bursts of decreasing amplitude on the IMF, with the smallest appearing at $t\sim1264.5 M_{0}$. This shows that the empirical mode decomposition of this signal effectively isolates junk radiation despite the presence of an underlying oscillatory pattern. This is possible because this trend is out of the frequency range isolated by the first IMF.

However, for many simulations, there are regions in the physical portion of the IMF that satisfy the junk radiation amplitude condition set by $f_{2}$ in Step \ref{stepB2.1}. As a result, the decay time is overestimated, and filters with higher $f$ value become necessary.

\subsection{Filter 3}\label{crit2}

We now consider simulations for which Filter~\ref{filter3} successfully resolves junk radiation, and compare its performance with that of Filter \ref{filter2}.

Simulation SXS:BBH:0433, shown in Fig.~\ref{fig:fig3}a, illustrates this case. We start by noting that the average amplitude of the IMF around the main burst ($15.5<t<281.0 M_{0}$) is $\sim1.29\times 10^{-6}$. Based on the values of $f_{2}$ and $f_{3}$ used by Filters \ref{filter2} and \ref{filter3}, IMF amplitudes greater $10^{-9}$ and $10^{-8}$ are considered indicative of junk radiation. With Filter~\ref{filter2}, the algorithm identifies at $t=3671.5M_{0}$ (not shown) a fluctuation with amplitude $1.34\times 10^{-9}$. With Filter~\ref{filter3}, however, it successfully identifies at $t_\mathrm{junk,EMD}= 1441M_{0}$ the last burst due to junk radiation, with an amplitude $1.41\times10^{-8}$. 

As shown above, Filter~\ref{filter3} deals with the cases for which Filter~\ref{filter2} fails. We emphasize, however, that both filters are necessary, each being appropriate to different sets of simulations. Applying Filter~\ref{filter3} to those that are best resolved by Filter~\ref{filter2} would lead to an underestimate of $t_\mathrm{junk}$. This is because the threshold values used by the third filter are less sensitive to the junk radiation features present in the simulations that are adequately dealt with by the second filter.

\subsection{Filter 4}\label{results_filter4}

Filter \ref{filter4} has the least strict criterion that still allows for the isolation of junk radiation. We found that increasing the value of the fraction $f$ shown in Step~\ref{step4.1} from 0.025 to $\gtrsim 0.03$ consistently leads to a significant underestimate of $t_\mathrm{junk}$. Accordingly, there is no point in introducing a fifth filter with a larger $f$ value. 

\begin{figure}
\centering
\includegraphics[width=0.5\textwidth]{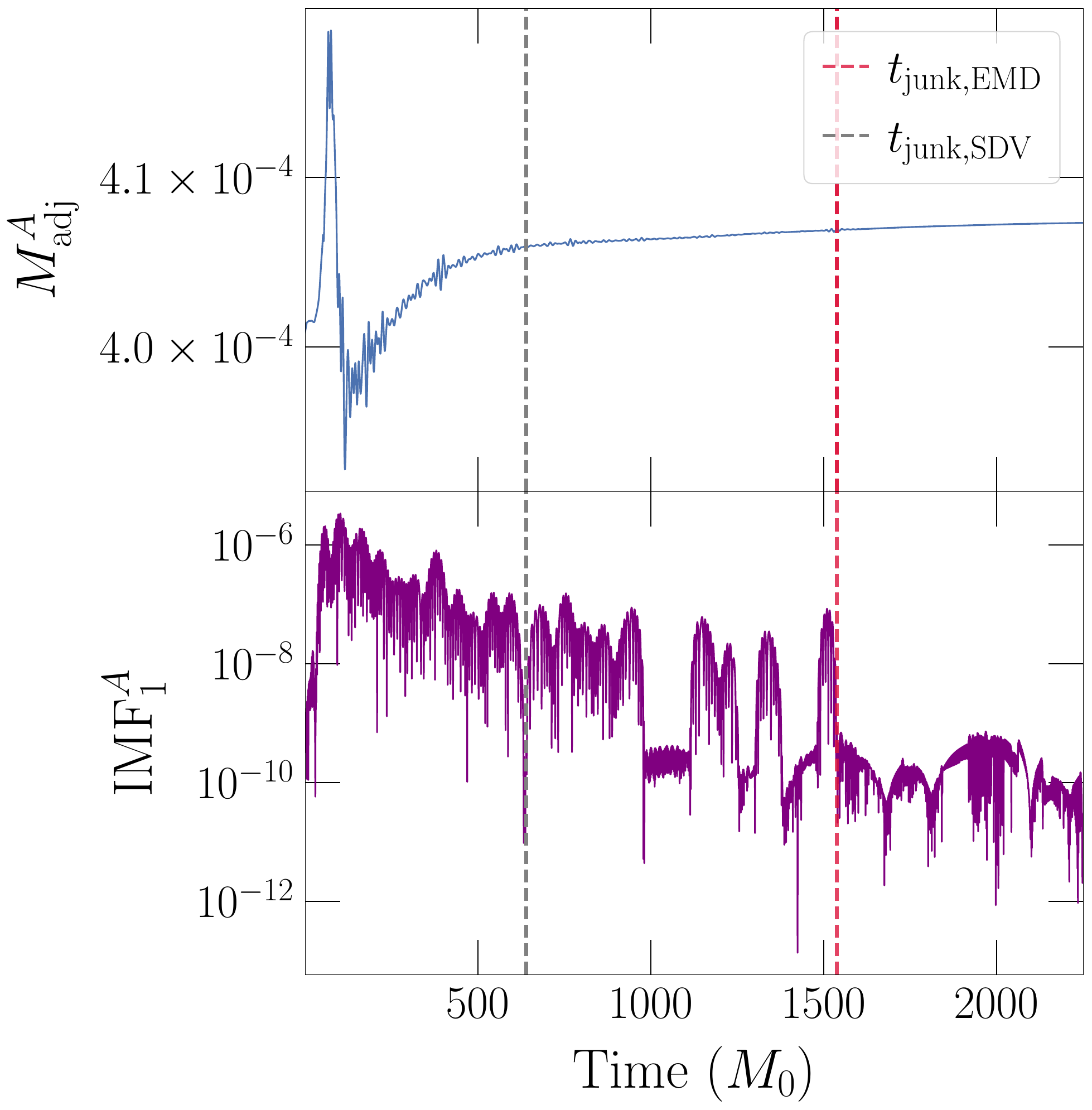}
\caption{The adjusted mass, with $M_\mathrm{offset} = 0.585$, and the first IMF of SXS:BBH:0504. The junk radiation is manifest as an initial pulse ($t\lesssim600 M_{0}$) followed by high-frequency fluctuations that become negligible at $t_\mathrm{junk,EMD} = 1538 M_{0}$. This is identified by Filter~\ref{filter4} of the EMD algorithm when the fraction $f_{4}=0.025$ is utilized. The grey dashed line shows the SDV estimate $t_\mathrm{junk,SDV} = 604 M_{0}$.}
\label{fig:fig5}
\end{figure}

We now compare the performance of Filters~\ref{filter3} and ~\ref{filter4} by considering the simulation SXS:BBH:0504, shown in Fig.~\ref{fig:fig5}. From the definitions given in Part A.\ref{A.3}, we find that $\texttt{mean\_about\_max}\sim7.00\times 10^{-7}$ for $13.5<t<312.5 M_{0}$, the window that characterizes the main burst. According to Filters ~\ref{filter3} and ~\ref{filter4}, a point on the IMF is considered junk radiation if its amplitude is $>7\times10^{-9}$ and $>1.75\times10^{-8}$. However, the average amplitude of the post-junk portion of the IMF ($t\gtrsim1540 M_{0}$) is $8.83\times10^{-8}$, causing Filter~\ref{filter3} to overestimate $t_\mathrm{junk}$. This is detected via Step~\ref{step3.2}, causing the algorithm to apply Filter~\ref{filter4} and find $t_\mathrm{junk,EMD}=1538M_{0}$.

We emphasize that there are no significant differences in the shape of the IMFs from Figs.~\ref{fig:fig3}a and ~\ref{fig:fig5}. However, as we have shown, variations in the ratio between the amplitude of the smallest burst from junk radiation and the average amplitude of the main burst makes distinct levels of resolution necessary.

\subsection{Eccentricity, precession and high spin}\label{exception}

The EMD approach performs especially well compared to the standard-deviation-based algorithm for systems in which the junk-free irreducible mass shows lower-frequency oscillations. These are typically associated with eccentric, precessing or high-spin runs.

\begin{figure}
\centering
\includegraphics[width=0.5\textwidth]
{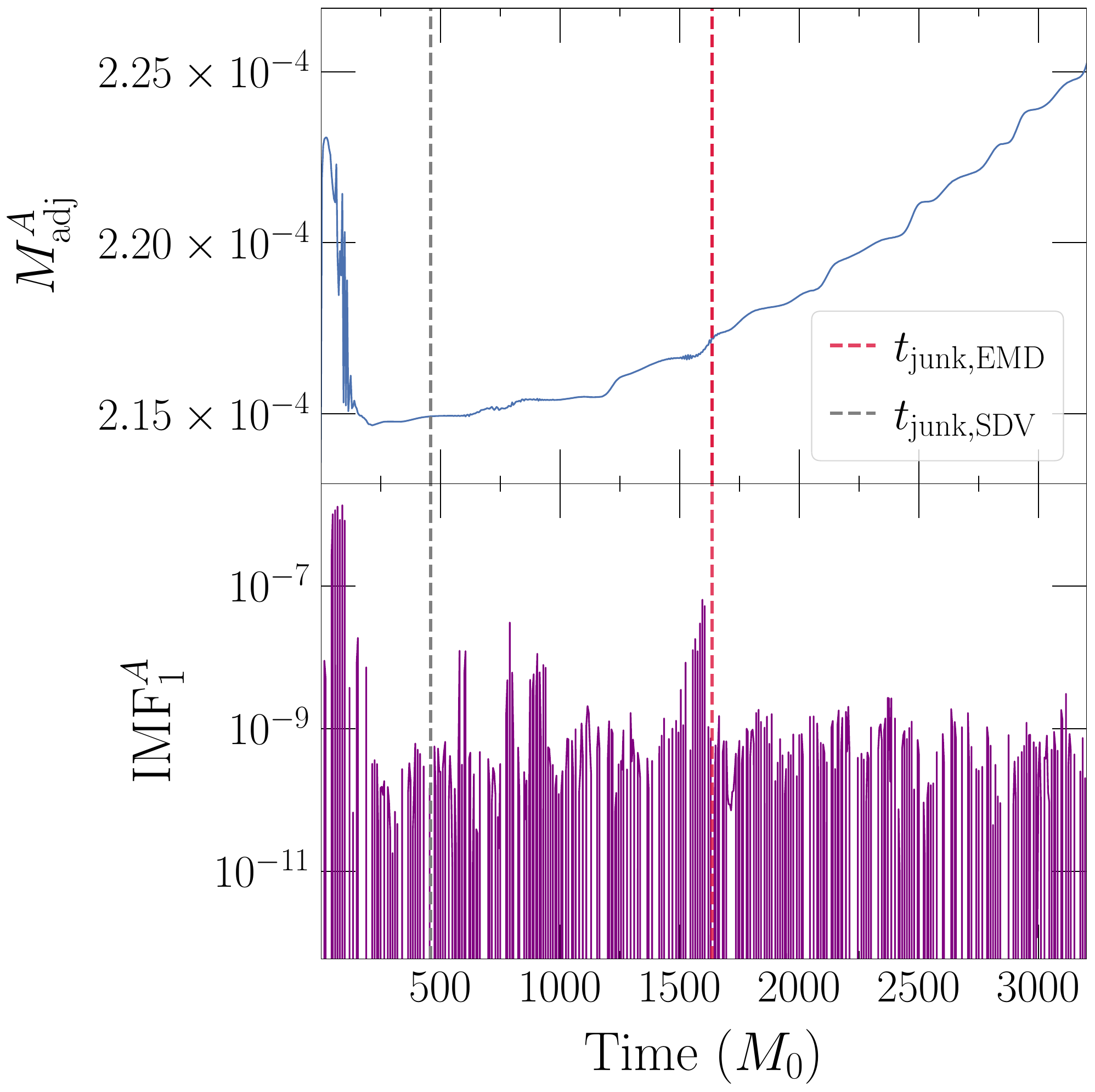}
\caption{The oscillatory pattern of SXS:BBH:3831$^{*}$. The red and grey dashed lines correspond to $t_\mathrm{junk,EMD} = 1720.5 M_{0}$ and $t_\mathrm{junk,SDV} = 460 M_{0}$. In the plot for the adjusted mass, which has $M_\mathrm{offset} = 0.447$, junk radiation can be observed at $500\lesssim t\lesssim 1720.5 M_{0}$ as high-frequency oscillations on top of the lower-frequency pattern that is characteristic of eccentric and precessing runs. These high-frequency contributions appear in the first IMF as localized bursts at $t\sim 1000M_{0}$ and $1500\lesssim t\lesssim 1720.5 M_{0}$. The physical oscillations in the irreducible mass are not captured by the remaining portion of the IMF ($t>t_\mathrm{junk,EMD}$).}
\label{fig:fig11}
\end{figure}

In Fig.~\ref{fig:fig11}, we illustrate the performance of our algorithm on the simulation SXS:BBH:3831$^{*}$, which has eccentricity $e=0.31$ and mass ratio $q=1$. 
In the plot for the adjusted mass, junk radiation is characterized by a nearly vertical feature at $t\lesssim 460 M_{0}$, followed by high-frequency fluctuations on top of the lower-frequency pattern of the physical evolution until $t\sim 1720.5 M_{0}$. These features are captured by the corresponding IMF at $t<460 M_{0}$, $t\sim1000 M_{0}$ and  $1500\lesssim t<1720.5 M_{0}$. This last burst is correctly identified by the algorithm, which returns $t_\mathrm{junk,EMD}= 1720.5 M_{0}$.

The effectiveness of our algorithm in simulations of this kind arises because the EMD decomposes the input data with junk radiation into components of different frequency ranges. The first IMF is therefore insensitive to the lower-frequency trend that underlies the high-frequency fluctuations from junk radiation. 

\subsection{Limitations}\label{limitations}

In this section we describe two cases in which the algorithm outlined in Sec.~\ref{sec:num impl} may fail to properly estimate $t_\mathrm{junk}$, along with possible solutions. 

First, the returned $t_\mathrm{junk}$ may not be accurate if the  IMF does not properly isolate junk radiation. This could occur if the frequencies that characterize junk radiation are not sufficiently distinct from those of the irreducible mass, potentially causing these different components to mix in the first IMF. If this issue affects the data of only one of the black holes, $t_\mathrm{junk}$ may still be correctly estimated at Step~\ref{stepC2.2} from Part~\ref{partC}. It is worth noting that we did not observe instances where the IMF for both black holes was affected by an incorrect tracking of junk radiation. 

\begin{figure*}
\begin{minipage}[b]{0.49\textwidth}
\centering
\includegraphics[width=\linewidth]{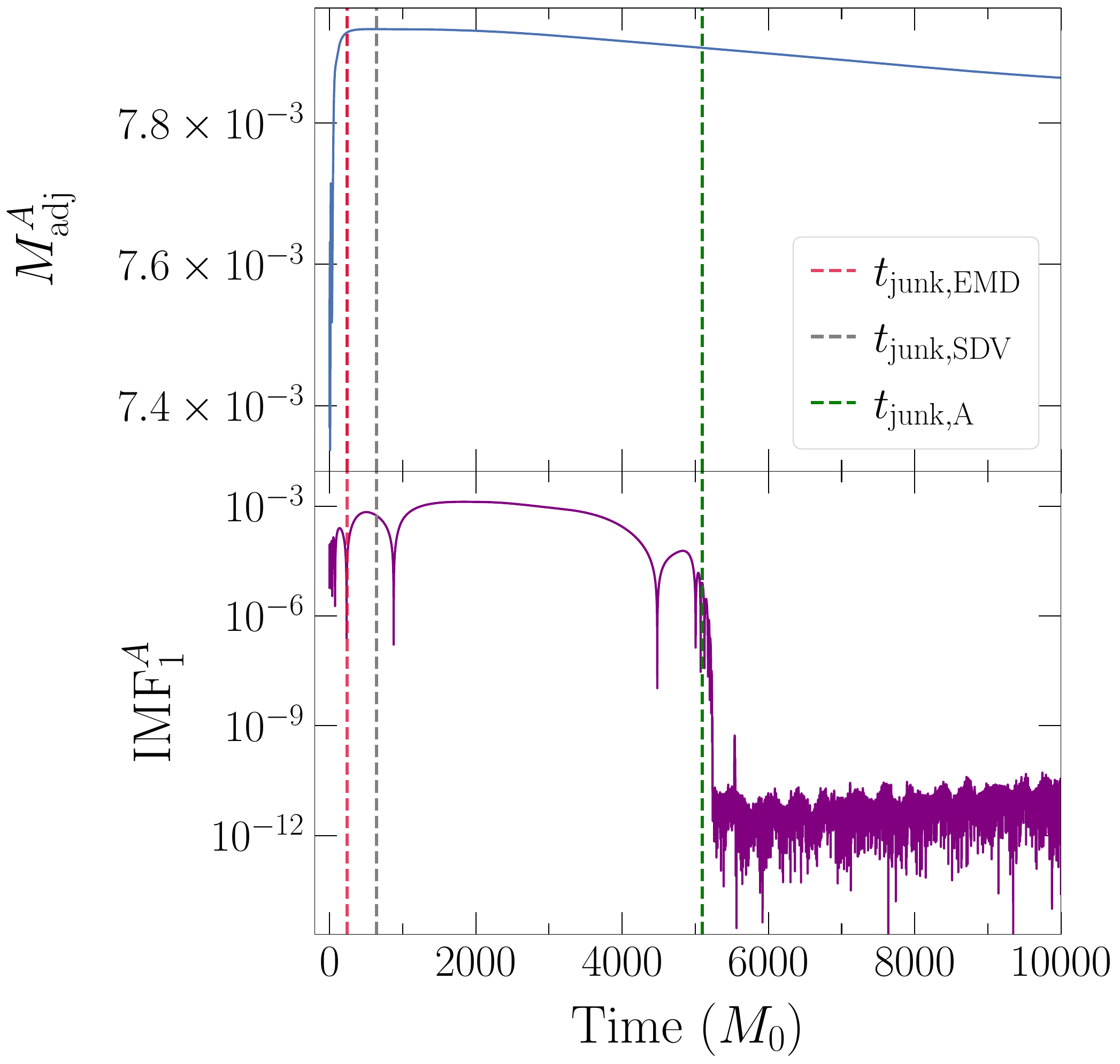}
\end{minipage}
\hfill
\begin{minipage}[b]{0.49\textwidth}
\includegraphics[width=\linewidth]{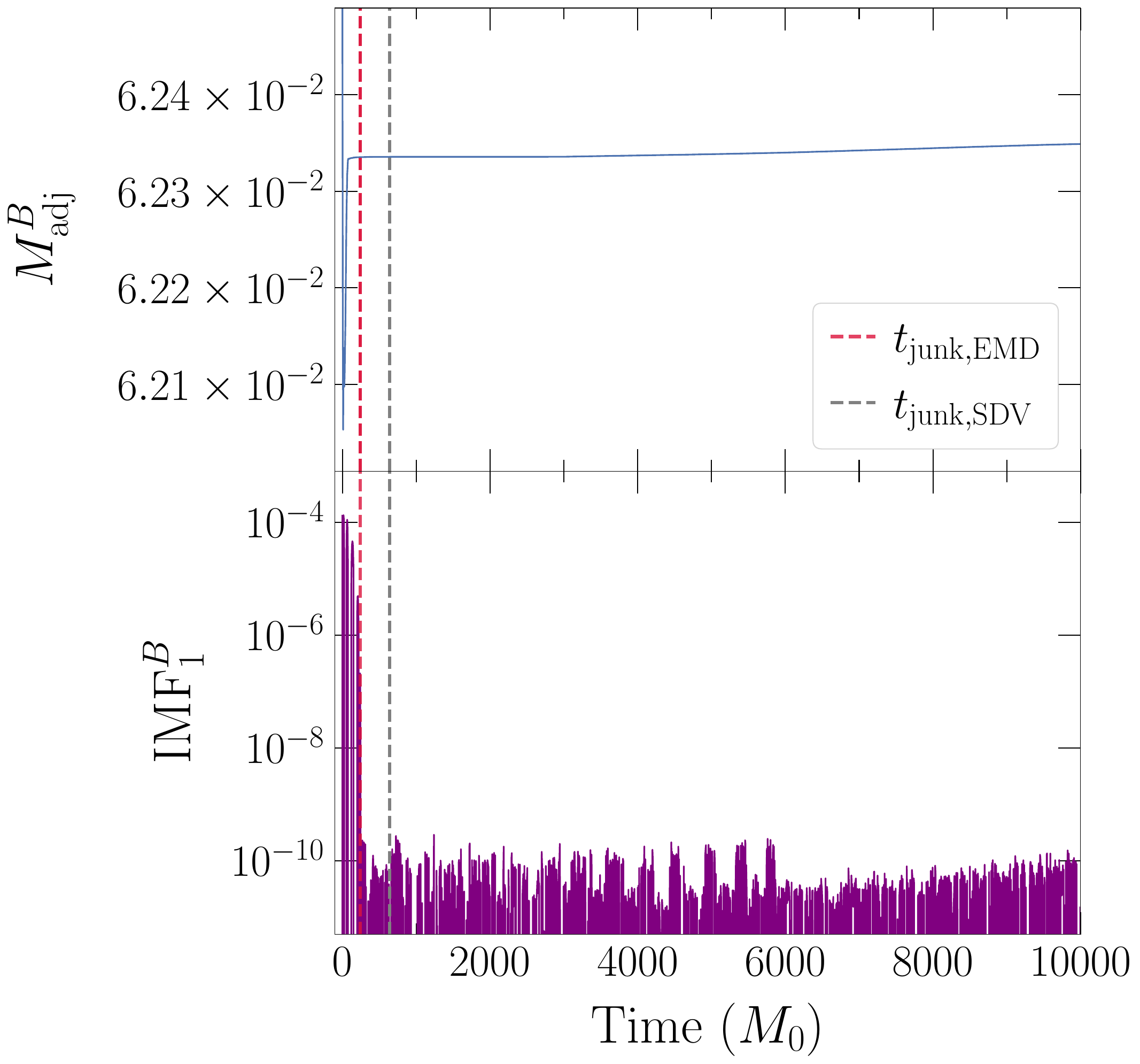}
\end{minipage}
\caption{The adjusted mass and first IMF for the primary (left/A) and secondary (right/B) black holes of SXS:BBH:382$2^{*}$. Here, the irreducible masses of the black holes are offset by $M_\mathrm{offset}^{A} = 0.93$ and $M_\mathrm{offset}^{B} = 0$. The EMD algorithm evaluated at the primary black hole yields $t_\mathrm{junk,A} = 5224 M_{0}$ (green/dashed). The result for the secondary black hole coincides with $t_\mathrm{junk,EMD} = 240 M_{0}$, as indicated by the red dashed line. For comparison, the grey dashed line represents the SDV-based estimate at  $t_\mathrm{junk,SDV} = 640 M_{0}$. Junk radiation is manifest for both black holes as the vertical feature at $t\lesssim600 M_{0}$. Compared to the IMF of the secondary black hole, that of the primary displays an additional oscillation at $640\lesssim t\lesssim5224 M_{0}$ that is unrelated to junk radiation. As a result, the estimated $t_\mathrm{junk}$ for both black holes differ by $\Delta t=4584M_{0}$. \label{fig:fig8}}
\end{figure*}

This case is exemplified in Fig.~\ref{fig:fig8}, which shows the adjusted mass and first IMF for both the primary (right) and secondary (left) black holes of the simulation SXS:BBH:382$2^{*}$. In the plot for the adjusted mass shown in Fig.~\ref{fig:fig8}a, junk radiation appears as the nearly vertical feature at $t\lesssim600 M_{0}$, causing the mass to fluctuate between $\sim7.3\times10^{-4}$ and $\sim 7.9\times 10^{-4}$. The IMF shows, in addition to the equivalent feature, one long-lived oscillation, unrelated to junk radiation, from $t \sim 640 M_{0}$ to $t\sim 5224 M_{0}$, with an amplitude of $\sim5\times10^{-4}$. The algorithm returns, as indicated by the green dashed line, $t_\mathrm{junk_{A},EMD}=5224 M_{0}$, which coincides with the end of this unrelated oscillation, but overestimates the actual junk radiation end time by over $4000 M_{0}$.  

In the plot for the adjusted mass shown in Fig.~\ref{fig:fig8}b, junk radiation is also displayed as a vertical feature at $t\lesssim600 M_{0}$, after which the signal stays approximately constant with an amplitude close to $6.23\times10^{-2}$. The first IMF captures this feature with an amplitude of $\sim 10^{-4}$, but does not display any significant oscillations afterwards. With this mapping, the algorithm finds $t_\mathrm{junk,EMD}= t_\mathrm{junk_{B},EMD}=240M_{0}$.

As this simulation shows, the IMFs from the two black holes may detect different patterns of oscillation. This is possibly due to the distribution of extrema in the input signal, which might affect the sifting that generates the IMFs (Sec.~\ref{emd}). Nonetheless, Part~\ref{partC} of the algorithm may detect this unusual behavior and still return an appropriate estimate for $t_\mathrm{junk}$.

A significant discrepancy ($\Delta t>3000M_{0}$) between $t_\mathrm{junk_{A}, EMD}$ and $t_\mathrm{junk_{B}, EMD}$, the values returned by the EMD algorithm evaluated at the two black holes, may indicate the presence of an unusual IMF. We observed this behavior for the simulations SXS:BBH:382$4^{*}$ and SXS:BBH:382$2^{*}$ only. These are extreme cases with mass ratio $q=15$, and the ambiguity in the determination of the junk radiation might indicate these runs are underresolved. Generally, it may also be that the IMFs from the two black holes differ by over $\Delta t>1000M_{0}$, but are individually correct. The divergence could reflect differences present in the time series, which naturally result in IMFs that identify different patterns. When this is the case, the discrepancy is not worrisome, since the largest returned value reflects real effects from junk radiation.

Second, we also identify simulations for which the default values for $T$ and $F$ used in Steps~\ref{stepB2.2},~\ref{step3.2} and~\ref{step4.2} causes the algorithm to overestimate $t_\mathrm{junk}$. We observe that this could occur if there are points in the IMF after junk radiation has decayed that meet the first, but not the second condition of Filters \ref{filter2}--\ref{filter4}.

\begin{figure}
\centering
    \includegraphics[width=\linewidth]{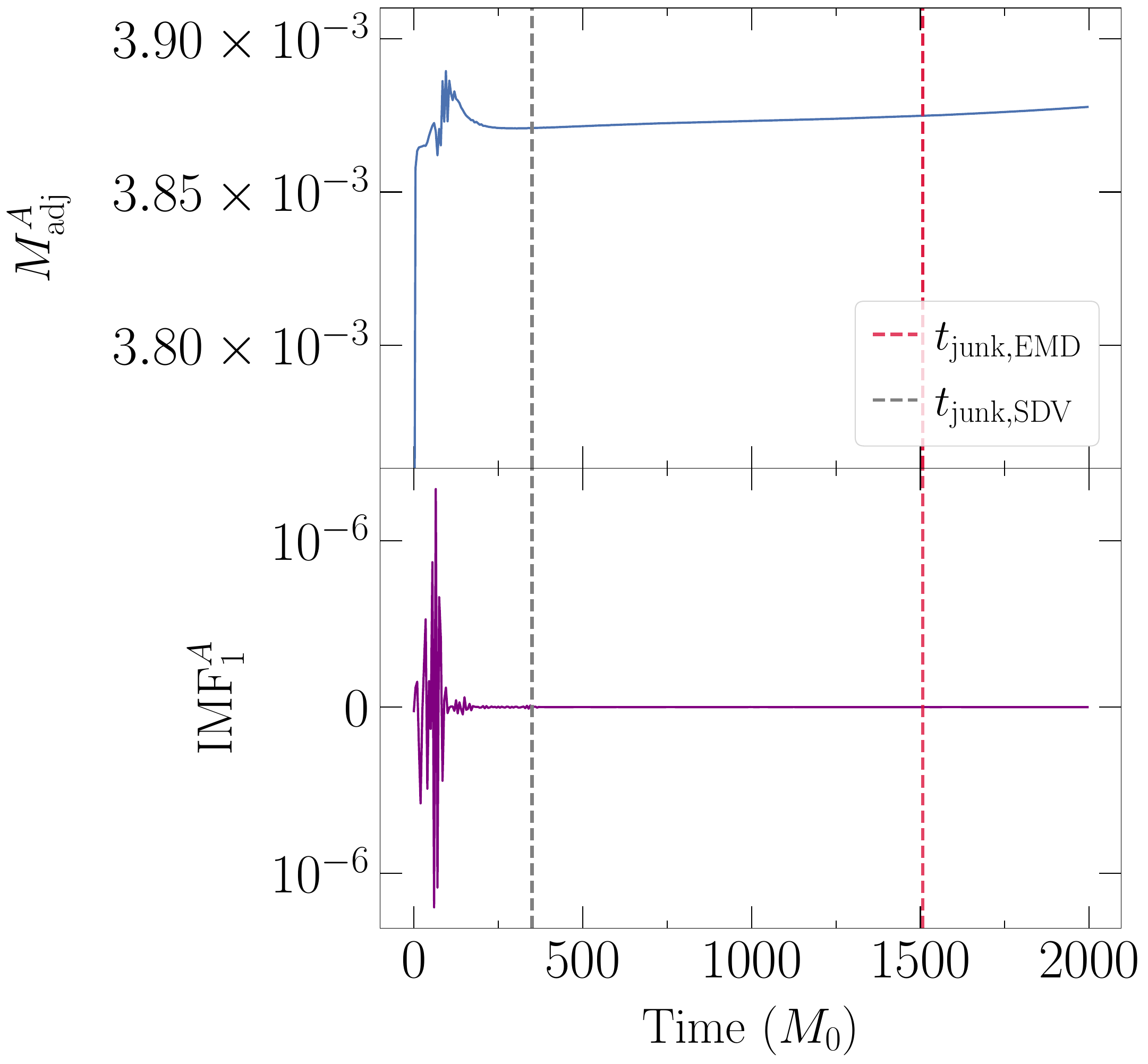}       
\caption{Filter \ref{filter3} applied to SXS:BBH:362$9^{*}$. In the plot for the adjusted irreducible mass (top/blue), junk radiation appears as a vertical feature at $t\lesssim 300M_{0}$ that decays into high-frequency fluctuations around the physical trend, subsiding at $t\sim 400M_{0}$. The irreducible mass is offset by $M_\mathrm{offset}=0.42$. The spurious bursts in the IMF (bottom/purple) become negligible in the same time interval. Evaluated with Filter \ref{filter3}, the EMD-algorithm (red/dashed) overestimates this timescale at $t_\mathrm{junk,EMD}=1506.5 M_{0}$. The SDV estimate is $t_\mathrm{junk,SDV}=350M_{0}$ (grey/dashed).}
\label{fig:fig10}
\end{figure}

In Fig.~\ref{fig:fig10}, we show the irreducible mass and IMF of SXS:BBH:3629 to exemplify this scenario. Following the initial feature at $t\lesssim 300M_{0}$, junk radiation is identified as a series of fluctuations around the physical trend that subside at $t\sim400M_{0}$. Similarly, the first IMF maps junk radiation as an initial nearly vertical feature followed by a main burst at $47.0\lesssim t\lesssim 233 M_{0}$ of amplitude $1.93\times10^{-6}$. The subsequent smaller bursts decay within $t\sim400 M_{0}$, leaving an IMF that is approximately constant. With Filter \ref{filter2}, the algorithm identifies a $2.08\times 10^{-9}$ fluctuation at $t=1506.5 M_{0}$, which corresponds to $53.2\%$ of  $t_\mathrm{end}=2831.0 M_{0}$, the latest time of the reduced time window. Although this point is evidently past the main component of junk radiation, it fails to meet condition \ref{stepB2.2}, and the algorithm returns $t_\mathrm{junk,EMD}=1506.5 M_{0}$. 

Similar cases can be fixed by changing the values of $F$ or $T$ used by Filters \ref{filter2}--\ref{filter4}, so as to trigger Filter \ref{filter4}. For the simulation from Fig.~\ref{fig:fig10} just examined, for example, one could trigger Filter \ref{filter4} by setting $F_{2} = F_{3} < 0.53$. Simulations that present this behavior are identified in Table~\ref{tab:simulation_ids2}. 
\\

\section{Conclusion} \label{sec:conclusion}

This paper describes an algorithm that implements the empirical mode decomposition to estimate the decay time of junk radiation in binary black hole simulations. We apply this method to the time series of irreducible masses of a binary black hole system, and find that the first IMF effectively isolates the high-frequency contributions from junk radiation. Based on this observation, we then establish threshold criteria that allow the user to determine the level of contamination they are willing to accept, and establish an upper bound on how much of the signal is expected to be contaminated by junk radiation. Upon applying this framework to the SXS catalog of BBH simulations, we find four categories based on the suitability of specific threshold criteria.  Aided by this classification, we create an algorithm that automatically applies the most appropriate criterion
for a given simulation. We then consider two cases that require slight modifications to the standard approach, which further shows the flexibility of the method. Lastly, we find that the EMD formulation is particularly suitable for the special case of simulations that are characterized by an oscillatory pattern.

In summary, the algorithm we introduce provides a framework to effectively estimate the decay time of junk radiation, so that the signal returned is free of unphysical fluctuations and initial quantities can be determined. We further note that the method described can be implemented regardless of the construction of the initial data; its effectiveness follows from the appropriate choice of threshold values. We also reiterate that we expect the procedure to be applicable to any time series that contains junk radiation with frequencies that are sufficiently high and distinct from its junk-free component. 

Future research could explore, for example, the application of our algorithm to binary neutron star simulations using data for their trajectories or central densities. Further studies could also be pursued to understand how the HHT formalism can be used to explore the long-term effects of junk radiation. This could illuminate ongoing debates about the changes junk radiation produces to the binary parameters, specifically to the spin and mass of the black holes. One may also investigate the extent to which the EMD may be applicable to gravitational-wave and numerical-relativity data to uncover physically meaningful patterns. 

\section*{Acknowledgements}
Computations for this work were performed with the Wheeler cluster at Caltech. This work was supported by the Sherman Fairchild Foundation and NSF Grants No. PHY-2309211, PHY-2309231, and OAC-2209656 at Caltech and PHY-2207342 and OAC-2209655 at Cornell.

\section*{Appendix A: Identification of Simulations}

The parameters of the binary systems analyzed in this paper and illustrated in Figs.~\ref{fig:fig1}-~\ref{fig:fig10} are shown in Table~\ref{tab:parameters}. 

The simulations described in the Secs.~\ref{results_filter1} and ~\ref{limitations} exhibit unusual behavior. We identify them in Tables
\ref{tab:simulation_ids1} and \ref{tab:simulation_ids2}. Those that are not yet publicly available as of the release of this paper are marked with an asterisk.

\begin{table}
\setlength{\tabcolsep}{1.8pt}
    \begin{tabular}{llllll}
      Fig.  &  Simulation ID & $M_{A}^\mathrm{init}/M_{B}^\mathrm{init}$ & $\rVert\vec{\chi}_{A}^\mathrm{init}\rVert$ &
      $\rVert\vec{\chi}_{B}^\mathrm{init}\rVert$ & $n_\mathrm{orbits}$ \\ 
       \hline\hline
       1 & SXS:BBH:1465 & 1.71 & 0.79 & 0.78 & 18\\
        2 & SXS:BBH:0692 & 1.67 & 0.80 & 0.80 & 18\\
        3 & SXS:BBH:0433 & 2.00  & 0.60 & 0.80 & 19 \\
        3 & SXS:BBH:1503 & 1.00 & 0.13 & 0.73 & 20 \\
        3 & SXS:BBH:1536 & 3.04 & 0.38 & 0.73 & 21 \\
        4 & SXS:BBH:1121 & 1.00 & 0.10 & 0.10 &  17 \\
        5 & SXS:BBH:278$1^{*}$ & 7.10 & 0.71 & $\sim$ 0 & 43\\
        6 & SXS:BBH:0504 & 1.44 & 0.26 & 0.41 & 19\\
        7 & SXS:BBH:383$1^{*}$ & 1.00 & 0.80 &  $\sim$ 0 & 17\\
        8 & SXS:BBH:382$2^{*}$ & 15.0 &  $\sim$ 0 &  $\sim$ 0 & 53\\
        9 & SXS:BBH:362$9^{*}$ & 1.00 & 0.90 & 0.90 & 13 \\      
    \end{tabular}
\caption{Parameters of simulations analyzed. The columns identify, in order, the figure number, simulation name, initial mass ratio, norm of the  dimensionless spin of the primary and secondary black holes at $t=0$, and the number of orbits.} 
\label{tab:parameters}
\end{table}

\begin{table}
\begin{ruledtabular}
\setlength{\tabcolsep}{25pt}
    \begin{tabular}{ll}
        SXS:BBH:0623&  SXS:BBH:363$6^{*}$ \\
        SXS:BBH:111$8^{*}$&  SXS:BBH:363$9^{*}$ \\
        SXS:BBH:112$1^{*}$& SXS:BBH:364$0^{*}$ \\
        SXS:BBH:1130& SXS:BBH:364$6^{*}$ \\
        SXS:BBH:1131& SXS:BBH:364$7^{*}$ \\ 
        SXS:BBH:1132& SXS:BBH:364$8^{*}$ \\
        SXS:BBH:1387&  SXS:BBH:365$3^{*}$ \\
        SXS:BBH:251$1^{*}$& SXS:BBH:365$4^{*}$\\ 
        SXS:BBH:329$5^{*}$& SXS:BBH:365$9^{*}$ \\ 
        SXS:BBH:363$1^{*}$& SXS:BBH:366$0^{*}$\\ 
    \end{tabular}
\end{ruledtabular}
\caption{Simulations that contain a small amount of junk radiation and that are best resolved by Filter~\ref{filter1}.
\label{tab:simulation_ids1}}
\end{table}

\begin{table}
\begin{ruledtabular}
\setlength{\tabcolsep}{25pt}
    \begin{tabular}{ll}
        SXS:BBH:0396& SXS:BBH:371$1^{*}$ \\
        SXS:BBH:061$2^{*}$& SXS:BBH:371$5^{*}$ \\
        SXS:BBH:249$8^{*}$& SXS:BBH:379$0^{*}$ \\
        SXS:BBH:362$9^{*}$& SXS:BBH:381$3^{*}$ \\  
    \end{tabular}
\end{ruledtabular}
\caption{Simulations that require a
modification of the numerical value of the parameters $T$
and $F$, as described in Sec.~\ref{limitations}.
\label{tab:simulation_ids2}}
\end{table}

\clearpage

\bibliography{bibliography}

\end{document}